\documentstyle[12pt,epsfig]{article}
\begin{document}
\title{Deep inelastic ratio $R=\sigma_{L}/\sigma_{T}$ and the possible 
existence of scalar partons in the nucleon
\thanks{Work supported in part by the KBN-Grant PB 2-P03B-065-10.}}
\author{Jan Bartelski, Wojciech Kr\'{o}likowski, Miros{\l}aw Kurzela\\
Institute of Theoretical Physics, Warsaw University,\\
Ho\.{z}a 69, 00-681 Warsaw, Poland. \\ }
\date{}
\maketitle
\vspace{1cm}
\begin{abstract}
\noindent We have performed several fits to find the ratio 
$R=\sigma_{L}/\sigma_{T}$ in different phenomenological models. 
Our fits seem to leave no room for possible admixture of scalar 
partons inside the nucleon.
\end{abstract}
\vspace{2cm}
\date{27 April 1998}
\newpage

The aim of this paper is to determine the possible admixture
of charged scalar partons in the nucleon. One is able to get such information
considering the ratio of virtual photon cross sections $\sigma_L/\sigma_T$ in 
the deep-inelastic scattering \cite{f}. This ratio R is proportional to the 
longitudinal structure function $F_L$,
\begin{equation}
R(x,Q^2)=\frac{\sigma_L}{\sigma_T}=\frac{F_{L}(x,Q^2)}{2xF_{1}(x,Q^2)}=
\frac{(1+\frac{4M_{p}^{2}x^2}{Q^2})F_2(x,Q^2)-2xF_1(x,Q^2)}{2xF_{1}(x,Q^2)}.
\end{equation}
In order to get $R(x,Q^2)$ 
from the data we have performed several phenomenological fits. We have used 
all deep-inelastic experimental points for  $Q^2>1 \mbox{\rm GeV}^2$ 
\cite{exp,c,NMC}, including the latest from the NMC Collaboration \cite{NMC}, 
which cover the kinematical range: 
$0.0045\leq x \leq 0.7$, $1 {\rm GeV}^2 \leq Q^2 \leq 70 {\rm GeV}^2$.

In the naive parton model (but with an inclusion of transverse momenta of 
nucleon constituents) one gets for the ratio R the formula \cite{f}
\begin{equation}
R(x,Q^2)=\frac{4m_T^2(x)}{Q^2},
\end{equation}
where $m_T^2$ is a quark transverse mass squared. Let us 
parametrize R as follows:
\begin{equation}
R(x,Q^2)=\frac{4M_T^2(x,Q^2)}{Q^2}.
\end{equation}
Then, in the naive parton model we get
\begin{equation}
M_T^2(x,Q^2)_{NQM}=m_q^2+k_T^2(x).
\end{equation}
The leading order QCD radiative corrections to this picture give 
in addition
\begin{equation}
M_{T}^{2}(x,Q^2)_{QCD}=\alpha_{s}(Q^2)Q^2f_{QCD}(x,Q^2),
\end{equation}
where $f_{QCD}$ is known if the parton (i.e. quark and gluon) distributions 
are available (see e.g. \cite{rob}). If we define \cite{mir1}
\begin{equation}
{\cal F}(x,Q^2)=x\sum_{a}e_{a}^2[q_{a}(x,Q^2)+\bar{q}_{a}(x,Q^2)],
\end{equation}
we obtain in LO QCD for electromagnetic interactions:
\begin{equation}
f_{QCD}(x,Q^2)=\frac{x^2}{3\pi}\frac{\int_{x}^{1}\frac{dy}{y^3}
{\cal F}(x,Q^2)}{{\cal F}(x,Q^2)}+\frac{x^2\sum_{a=1}^{n_F}e_{a}^2}{2\pi}
\frac{\int_{x}^{1}\frac{dy}{y^3}(y-x)G(x,Q^2)}{{\cal F}(x,Q^2)},
\end{equation}
where $G(x,Q^2)$ is the gluon distribution inside a nucleon and we sum over 
$n_F$ quark flavors.

Because the contribution to $M_{T}^2$, coming from QCD radiative corrections, 
is of order $\alpha_sQ^2$, it is usually bigger that the term calculated in 
the parton model which is of  order $M_N^2$ ($M_N$ stands for nucleon mass).

If we take into account the possible existence of charged scalar partons 
inside the nucleon, we get an additional term
\begin{equation}
M_T^2(x,Q^2)_{SC}=Q^2f_{SC}(x,Q^2),
\end{equation}
where the unknown function $f_{SC}$ is connected to the function $\gamma (x)$
introduced in ref.\cite{f}:

\begin{equation}
f_{SC}(x)=\frac{1-\gamma (x)}{4 \gamma (x)}.
\end{equation}
Hence, because of different $Q^2$ dependence of all three terms (see 
eqs.(4,5,8)) it is, in principle,
possible to extract an information about scalar parton contribution.

A priori, scalar constituents might enter into the nucleon structure 
in different ways. Let us mention two options corresponding to two 
possible variants in the origin of these scalars.

(i) They could appear in an extended nucleon sea as squark-antisquark pairs
$\tilde{q}\tilde{\bar{q}}$ or, alternatively, as some other quarklike 
scalar-antiscalar pairs $y\bar{y}$. 
In particular, the latter partons $y$ might be the quarklike scalars named 
recently "yukawions" by one of us \cite{kr1}.

(ii) In a composite model of  {\em u} and {\em d} quarks interpreted as bound
states of a spin-1/2 preon {\em U} or {\em D} and  a spin-0 preon $\phi$, the 
scalars $\phi$ could manifest in a generalized nucleon sea as 
$\phi$$\bar{\phi}$ pairs, and/or in a generalized nucleon valence fraction as 
single $\phi$ constituent arising in a dissociation process {\em u},{\em d} 
$\rightarrow$ {\em U},{\em D} + $\phi$ \cite{kr2}. In particular, 
the spin-1/2 preon {\em U}, {\em D} might carry the electric charge 1 or 0, 
respectively, the lepton number L = - 1 and no color, while the spin-0 preon 
$\phi$  would be a leptoquark with the electric charge -1/3 and the baryon 
and lepton numbers B=1/3 and L=1 \cite{kr3}. Note that such {\em U} and 
{\em D} preons, together with the $\nu_e$ and $e^{-}$ leptons treated as 
elementary, would form an anomaly-free set of fundamental fermions 
(of the first family).

It is intuitively clear that, in the case of relatively heavy scalars, their 
role in deep inelastic scattering off the nucleon should increase with $Q^2$. 
Though it does not prove to be true so far, this is especially suggestive in 
the case of option (ii) based on a picture of quarks {\em u} and {\em d} 
dissociating into their preons.

In our analysis we have repeated, first of all, the phenomenological fit 
presented in ref.\cite{whit} with inclusion of new data from the NMC 
collaboration \cite{NMC}. The parametrization in such a fit is:
\begin{equation}
R(x,Q^2)=\frac{b_1}{\log(\frac{Q^2}{0.04})}[1+\frac{12Q^2}{Q^2+1}
\frac{0.125^2}{0.125^2+x^2}]+\frac{b_2}{Q^2}+\frac{b_3}{Q^4+0.3^2},
\end{equation}
where $Q^2$ is in the ${\rm GeV}^2$ units. The first term simulates the LO 
QCD prediction, the second and the third mimic twist effects. 
However, such parametrization does not take into account  that ratio R can be 
different for different nucleon targets. On the other hand, present 
experimental results suggest $R_d \simeq R_p$ \cite{exp}. 

The parameters in our first fit, giving $\chi^2$/d.o.f. = 143/174  
$\simeq$ 0.82, are (in parenthesis we quote parameters obtained in 
\cite{whit}):
\begin{equation}
\begin{array}{ll}
b_{1}=\hspace*{0.293cm} 0.041 \pm 0.006&\hspace*{0.293cm}(0.0635),\\
b_{2}=\hspace*{0.293cm} 0.592 \pm 0.009&\hspace*{0.293cm}(0.5747),\\
b_{3}=-0.331 \pm 0.010&(-0.3534).\\
\end{array}
\end{equation}
Next, we have tried to incorporate the term which comes from hypothetical, 
electromagnetic active, scalar partons. We have modeled the unknown function 
$f_{SC}(x,Q^2)$ (eq.(7)) very simply, namely as a combination of only two 
functions ${\cal F}(x,Q^2)$ (eq.(6)) and ${\cal F}_{val}(x,Q^2)$, where
\begin {equation}
{\cal F}_{val}(x,Q^2)=x\sum_{a}e_{a}^2[q_{a}(x,Q^2)-\bar{q}_{a}(x,Q^2)]
\end{equation}
describes valence quarks, provided the quark and antiquark distribution in 
the sea are identical.
We propose for scalar parton contribution the following ansatz:
\begin{equation}
(M_T^2)_{SC}=\frac{1}{4}pQ^2\frac{\lambda{\cal F}_{val}(x,Q^2)+
\mu{\cal F}(x,Q^2)}{{\cal F}(x,Q^2)},
\end{equation}
where we choose $\lambda$=1 and $\mu=0$, what corresponds to the conjecture 
that scalar partons are distributed similarly to valence quarks. On the other hand, it turns 
out that such a choice, where $\em p$ is the only parameter, gives the 
optimal fit to data. For such a parametrization the $\chi^2$ value is not 
changed ($\chi^2 \simeq 143$) and leads to the value 
$\chi^2$/d.o.f. = 143/173 $\simeq$ 0.83 that is worse than in the previous 
case. The parameter $\em p$ which plays roughly the role of 
probability for finding scalar parton inside a nucleon (see e.g. \cite{f}) is 
$p=0.14\pm 0.48 \%$ i. e., consistent with zero. For quarks and antiquarks 
we used parton distributions proposed by the authors of ref.\cite{grv}.

The second type of model for the ratio R discussed here is the leading order 
QCD formula (for four flavors) with inclusion of simple parametrization of the twist effects (the third 
and the fourth term in the following expression):
\begin{eqnarray}
R(x,Q^2)&=&\frac{4x^2}{3\pi}\alpha_s(Q^2) \int_{x}^{1}\frac{dy}{y^3}
{\cal F}(x,Q^2)+ \\ \nonumber
&+&\frac{20x^2}{9\pi}\int_{x}^{1}\frac{dy}{y^3}(y-x)G(x,Q^2)+
\frac{4m_T^2}{Q^2}+\frac{W}{Q^4}.
\end{eqnarray}
Here, we consider $m_T^2$ and $W$ as two parameters (i.e., constants 
independent of $x$; for $m_T^2$ we follow a conjecture of refs.\cite{k2,mir1}). 

Taking the quark and gluon distribution 
from GRV fit \cite{grv}, we get $\chi^2$/d.o.f. = 141/153  $\simeq$ 0.92, and
\begin{equation}
\begin{array}{lll}
m_T^2&=&\hspace*{0.293cm}( 0.310\pm 6\; {\rm MeV})^2 ,\\ \nonumber
W&=&-(0.69 \pm 0.01\; {\rm GeV})^4.\\ 
\end{array}
\end{equation}
Performing the last fit we have not included the neutrino data from the {CDHSW} 
collaboration \cite{c}. If we add to such a fit a new term, which comes from 
the possible admixture of scalar constituents inside a nucleon (see eq.(13)) 
we get a similar values for $\chi^2$ and for parameters $m_T^2$ and $W$, 
whereas the parameter $p$ is consistent with zero 
($p\simeq 0.25 \pm 0.50 \% $). The best fit is obtained by analyzing 
the data for $Q^2 \geq 10\; {\rm GeV}^2$ only (this enables us to get rid 
of unknown twist contributions).
Then $\chi^2$/d.o.f. = 40/58  $\simeq$ 0.69 and
\begin{equation}
\begin{array}{lll}
m_T^2&=&\hspace*{0.293cm}( 0.39\; {\rm GeV})^2 ,\\ \nonumber
W&=&-(1.22\; {\rm GeV})^4,\\ 
\end{array}
\end{equation}
whereas $p$ is still consistent with zero.

In figures 1 and 2 we compare our fits, eqs.(10) and (14) to experimental 
data for definite $Q^2$ ($Q^2 =10\; {\rm GeV}^2$ and $Q^2 =45\; 
{\rm GeV}^2$). 
One sees that  
the difference between  them is visible only for small $x$. In figure 3 the 
curves for the second fit, eq.(14), calculated for different targets (proton 
and deuteron), are compared with the experimental points for 
$Q^2 =10 \; {\rm GeV}^2$.
\vspace{6cm}
\begin{figure}[thb]
\vskip 0cm\relax\noindent\hskip 0.5cm
       \relax{\includegraphics{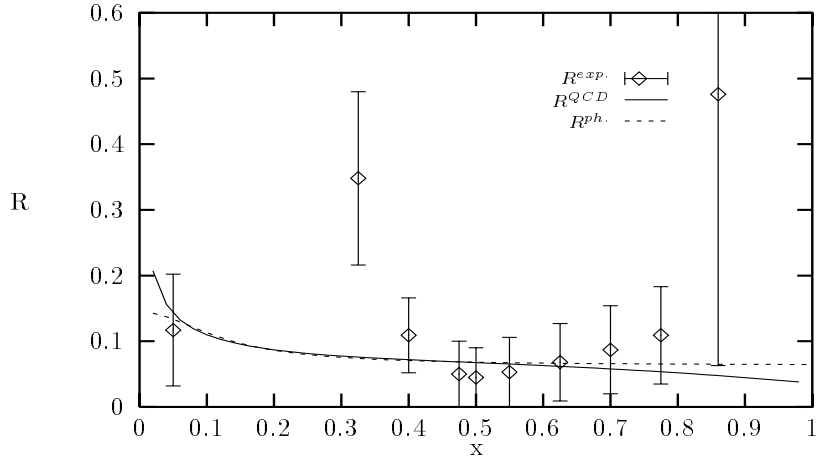}}
\caption{\em The comparison of formulae: eq.(10) with parameters (11) and
eq.(14) with parameters (15) ($R^{ph.}$ and $R^{QCD}$, respectively) with
the experimental data for $Q^2=10\;GeV^2$ \protect\cite{exp,c,NMC}. }
\label{fig1}
\end{figure}
\vspace{6cm}
\begin{figure}[hbt]
\vskip 0cm\relax\noindent\hskip 0.5cm
       \relax{\includegraphics{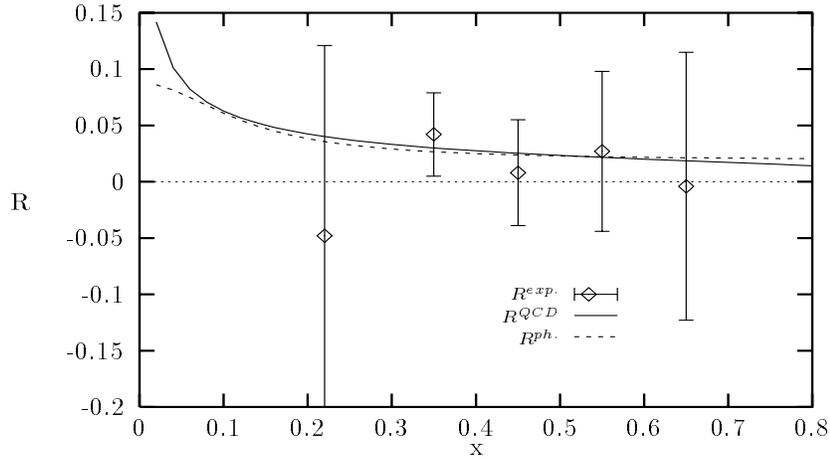}}
\caption{\em The comparison of formulae: eq.(10) with parameters (11) and
eq.(14) with parameters (15) ($R^{ph.}$ and $R^{QCD}$, respectively) with
the experimental data for $Q^2=45\;GeV^2$  \protect\cite{exp,c,NMC}. }
\label{fig2}
\end{figure}

In conclusion, we have analyzed two different models for ratio $R$, one 
phenomenological and one inspired by QCD with twist corrections added, 
getting no sign of existence of scalar constituents inside the nucleon. 
\vspace{6cm}
\begin{figure}[htb]
\vskip 0cm\relax\noindent\hskip 0.5cm
       \relax{\includegraphics{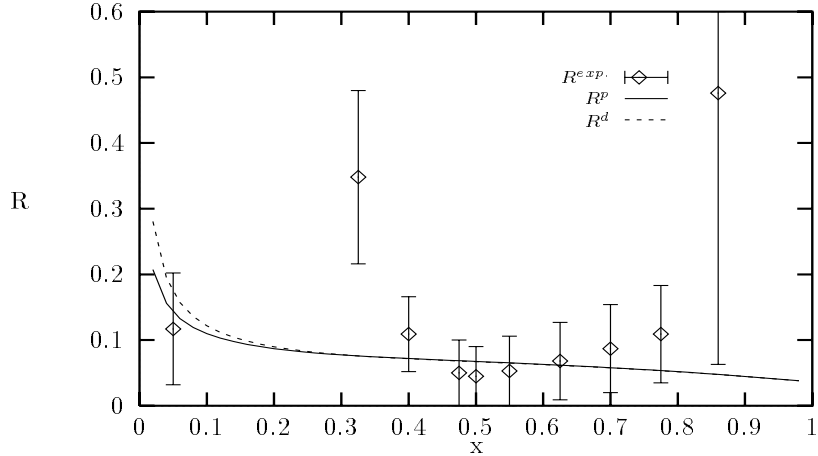}}
\caption{\em The comparison of formulae eq.(14), calculated for different
targets (proton: $R^p$ and deuteron: $R^d$) with the experimental data
for $Q^2=10\;GeV^2$  \protect\cite{exp,c,NMC}. }
\label{fig3}
\end{figure}

% A useful Journal macro
\def\jou#1#2#3#4{{#1} {\bf #2}, #3 (#4)}

% Some useful journal names
\def\NCA{{Nuovo Cimento} {\bf A}}
\def\NIM{Nucl. Instrum. Methods}
\def\NIMA{{Nucl. Instrum. Methods} {\bf A}}
\def\NPB{{Nucl. Phys.} {\bf B}}
\def\PLB{{Phys. Lett.}  {\bf B}}
\def\PRL{{Phys. Rev. Lett.}}
\def\PRD{{Phys. Rev.} {\bf D}}
\def\ZPC{{Z. Phys.} {\bf C}}
\def\APPB{{Acta Phys. Pol.} {\bf B}}

\end{document}